\documentclass[nofootinbib,onecolumn]{revtex4}
\usepackage{graphicx}
\usepackage{subfig}
\usepackage{amssymb}
\usepackage{amsmath}
\usepackage{bm}
\usepackage{hyperref}
\usepackage{verbatim}

\begin{document}
\title{Some effects of non-conservative gravity on cosmic string configurations}
\author{E. A. F. Bragan\c{c}a}
\email{eduardo.braganca@ufes.br}
\author{Thiago R. P. Caram\^es}
\email{trpcarames@gmail.com}
\affiliation{N\'ucleo Cosmo-ufes e Departamento de F\'isica, CCE, Universidade Federal do Esp\'irito Santo (UFES)\\
Vit\'oria-ES, Brazil}
\author{J\'ulio C. Fabris}
\email{julio.fabris@cosmo-ufes.org}
\affiliation{N\'ucleo Cosmo-ufes e Departamento de F\'isica, CCE, Universidade Federal do Esp\'irito Santo (UFES)\\
Vit\'oria-ES, Brazil}
\affiliation{National Research Nuclear University “MEPhI”\\
Kashirskoe sh.  31, Moscow 115409, Russia}

\author{Ant\^onio de P\'adua Santos}
\email{ antonio.padua@ufrpe.br}
\affiliation{Departamento de F\'isica, Universidade Federal Rural de Pernambuco (UFRPE)\\
               Rua Dom Manuel de Medeiros S/N,  Dois Irm\~aos - CEP 52.171-900 Recife-PE, Brazil}

\begin{abstract} 
We investigate some consequences of a specific non-conservation of the energy-momentum tensor on the physics of certain cosmic string configurations. This non-conservation is induced by a new gravitational theory recently introduced as an attempt to incorporate dissipative systems in the gravity description\footnote{Since a non-trivial conservation law is often achieved within many different alternative theories, maybe lacks exactness in using ``non-conservative gravity" to refer to the present gravitational theory we are studying here. However, we shall keep this nomenclature in order to avoid conflicts with the terminology already established in the literature regarding this issue.}. This model of gravity is endowed with a variational principle, where an action-dependence on the geometric sector is assumed. For the Abelian-Higgs string we derive the dynamical equations and provide the numerical solutions describing the behavior of the Higgs and gauge fields as well as the profile of the metric functions for different cases of the parameter that characterises the model. Additionally, from this numerical approach we find how the inner structure of the cosmic string is affected by this underlying gravitational theory, by analyzing the deviations both on the mass per unit length and the deficit angle of the string. Next, we proceed with an analytical treatment obtaining the solution close to the string axis and also the corresponding vacuum solution. Our analytical results are also complemented with the study of a thick cosmic string model, where the defect is endowed with finite core. In this case we have obtained the proper interior solution which joined together with the vacuum metric outside provides an exact solution for the exterior geometry which generalizes previous results and leads to a decreasing of the mass per unit length of the cosmic string.          
\end{abstract}

\maketitle
\noindent

\section{Introduction}
The formulation of a least action principle suitable to describe dissipative systems is object of a longstanding discussion within the theoretical physics. In the context of classical mechanics one usually deals with it by means of the dissipation Rayleigh function which comes up as a correction at the level of Euler-Lagrange equations \cite{goldstein}. In 1930, G. Herglotz succeeded to construct an extension of classical mechanics in which the dissipative contributions are able to appear already at the action level \cite{herglotz}, by assuming a dependence of the lagrangian function upon the action itself. Only recently a covariant version of the Herglotz theory emerged in \cite{lazo}, where the authors set out the foundations of a new gravitational theory whose scope encompasses the description of the dissipative systems. One of the immediate consequences of it is that the theory will bring a breaking of the diffeomorphism invariance which shall be translated into a non-conservation of the energy-momentum tensor.
The cosmology resulting of this modified gravity was recently investigated in \cite{fabris}, where the authors found out an interesting equivalence between the corresponding background dynamics of such a cosmology with that one of a bulk viscous universe. At perturbative level, they have shown that the dynamics of the matter perturbations indicates that some drawbacks faced by the bulk viscous model may be alleviated within such a non-conservative framework.

The potential appearence of topological defects in the earliest instants of the universe's evolution is a significant possibility within the context of grand unified theories (GUT) \cite{kibble,VS,hindmarsh,vilenkin}. Theoretically, it is expected that such objects contain a huge amount of energy, far greater than the scales achievable at the known accelerators, indicating that their detection would allow us to access high energy scales not yet known, thus leading to a proper comprehension of key aspects of the beyond standard model physics. In this picture, topological defects are relics left behind after symmetry breaking processes triggered by the many phase transitions experienced by the early universe during its expansion and progressive cooling down. The nature of the symmetry breaking determines the specific type of topological defect to be formed. Cosmic string is a line-like defect, resulting from a breaking of an axial symmetry, which is somewhat analogue to vortex lines in superfluids and superconductors, as well as dislocation-like imperfections in crystal structures. The existence of these objects are also especulated in the context of string theory where cosmic strings are considered as fundamental strings which stretched to a cosmological scale \cite{witten}. Besides, one expects cosmic strings may be formed in some inflationary models, as in the brane inflation scenario \cite{sarangi}. Possible consequences in the cosmological or astrophysical contexts of the existence of cosmic strings have been already extensively investigated. In the past, one used to think of cosmic strings as a possible provider of the generating mechanism of the primordial density fluctuations, competing with the inflationary model on this same job. However, further observations of the cosmic microwave background (CMB) supported the latter as the prevailing source of such a fluctuations \cite{wmap,south}, which was corroborated by the most recent observations \cite{planck1,planck2}. Despite of that, it is well known that one possible observational window for cosmic strings is the gravitational-wave radiation they can produce. This can occur in several processes involving cosmic strings loops as for examples decay, oscillations or self-intersections of these loops \cite{VS}. So, the revolution in astrophysics and cosmology inaugurated by the first direct detection of gravitational waves by the advanced LIGO/Virgo interferometers shall represent to the cosmic strings issue a new chance for its so expected detection (see \cite{PLB} for a recent discussion).  

All the known gravitational effects of cosmic strings shall be obviously affected by the underlying gravity theory in which they are studied. The Lazo {\it et al} modified gravity is a theory where dissipative effects emerge naturally from the least action principle, implying in a departure from the standard conservation of the energy-momentum tensor. It is expected that a cosmic string feels such a deviation through a change in its mass per unit length or linear energy density (which is equal to its tension if we consider the string a system with boost-invariance along its axis), what would affect the main observable quantity of a cosmic string setup. In this vein, our purpose in this work is to investigate such a influence of this modified gravity on the main properties of some cosmic strings configurations. 

The paper is organized as follows: in the next section the aforementioned gravitational theory is briefly introduced with the corresponding equations that dictate its dynamics. In the section III we get in touch with the Abelian-Higgs model, the first cosmic string scenario that we shall explore in this work. We derive the explicit form of the respective field equations and show that the non-conservative aspect of the theory may reduce its dynamical equations to those ones describing the Bogomolnyi-Prasad-Sommerfeld (BPS) limit \cite{Bogomol'nyi,PS:1975}. In the section IV we present the full numerical solutions for this problem, showing the impact of the model's parameter on the profile of the metric function along with the behaviour of the gauge and the Higgs fields. In the section V, we report the analytical results obtained in this work, covering both approximate and exact cases. In our approximative approach we find how this extended gravity shapes both the geometry and the form of the matter fields close to the cosmic string axis, revealing consistency with the full numerical treatment when distances near the string are taken. On the other hand, the exact cases comprise i) the vacuum solution, which shall also correspond to the asymptotic behaviour of the Abelian-Higgs string far away from the defect and ii) the generalization of the Hiscock-Gott solution \cite{Hisc,gott}, which means an exterior solution for a cosmic string with finite thickness which has a uniform energy density contained inside it. Once obtained such a solution, we discuss how does it affect the structure of the string by investigating possible variations in its mass per unit length. In the section VI we shall present our concluding remarks about this study as well as some possible extensions to be better studied in a future work.       

\section{The theory}
Originally an action-dependence lagrangian was used by Herglotz to deal with the principle of least action for dissipative systems in the classical mechanics context \cite{herglotz}. Recently, Lazo {\it et al} generalized this theory, translating it to a covariant language \cite{lazo}. The authors propose a model of gravity described by the following lagrangian
\begin{eqnarray}
\label{ADL}
{\cal L} = \sqrt{-g}(R - \lambda_{\mu}s^\mu) + {\cal L}_m,
\end{eqnarray}
with $s^\mu$ being an action-density field (which disappears during the variation of the action, so it does not manifest in the field equations) and $\lambda_\mu$ is a parameter encoding the emerging of ``geometric'' dissipative effects in the gravity and the deviation from General Relativity (GR). A proper justification for the choice (\ref{ADL}) can be found in the references \cite{lazo} and \cite{fabris}. In the Herglotz formalism the friction forces (with linear dependence in the velocity) arises in the equations of motion when one assumes a Lagrangian with linear dependence in the action. In this vein, in \cite{lazo} the authors argue that (\ref{ADL}) is the ``best" possible analogy to that linear-in-action dependence of the Lagrangian in a covariant framework. In Ref. \cite{fabris} the authors provide a more detailed discussion about the motivations behind this choice.      

Let us remark that according to the original formulation for this model, such an action-dependence is only upon the Einstein-Hilbert action, with no dependence on the matter action being taken into account. This means that the novel effects arising from this gravitational theory are of purely geometric origin. This coupling four-vector $\lambda^{\mu}$ may be in general a running variable, although we shall assume throughout this work the simplest case where its components are constant. 

The set of field equations of the theory proposed in \cite{lazo} can be written as follows
given by
\begin{eqnarray}
\label{feqs}
R_{\mu\nu}+K_{\mu\nu}=\frac{8\pi G}{c^4}(T_{\mu\nu}-\frac{1}{2}g_{\mu\nu}T),
\label{03}
\end{eqnarray}
where $K_{\mu\nu}$ is defined:
\begin{equation}
\label{kmn}
K_{\mu\nu}=\lambda_{\alpha}\Gamma_{\mu\nu}^\alpha-\frac{1}{2}(\lambda_\nu \Gamma_{\mu\alpha}^\alpha +
\lambda_\mu\Gamma_{\nu\alpha}^\alpha).
\end{equation}
We refer the readers to the Ref. \cite{lazo}, where a step-by-step derivation of the equations (\ref{feqs}) is given. In (\ref{kmn}) the parameter $\lambda^{\mu}$ is a four-vector encoding the non-conservative nature of the theory. The set of equations is also complemented with the modified conservation law given by
\begin{equation}
\label{nclaw}
\nabla_\mu(K_\nu ^\mu-\frac{1}{2}\delta_\nu^\mu K)=\frac{8 \pi G}{c^4}\nabla_{\mu}T^{\mu}_{\nu}.
\end{equation}
From now on we shall consider $c=1$.

It is easy to check that Eq. (\ref{nclaw}) when combined with (\ref{feqs}), ensures the fullfilment of the Bianchi identity. Such a equation is a natural consequence of the breaking of the diffeomorphism invariance experienced by the theory, essentially due to the presence of the four-vector $\lambda_{\mu}$, which allows for a preferred direction in the spacetime. However, notice that locally the variation of energy-momentum tensor is given in terms of the ordinary derivative, so it behaves locally in the same way as in GR. Nonetheless, when the curved spacetime geometry is important, one expects some deviation from GR. Such a deviation shall have physical consequences, which can only be examined when specific gravitational problems are addressed. One example of this, as mentioned in the Introduction, is the bulk viscous model which appears at background level, when this theory is used in the cosmological context \cite{fabris}. In \cite{lazo1} the authors use this same formalism of action-dependent Lagrangians to a wider class of physical systems, extending their approach to contexts different from the gravitational one.

\section{Abelian-Higgs strings}
We are interested in studying cosmic string configurations and its gravitational effects within such a extended theory of gravity. We start with the Abelian-Higgs string which is described by the following Lagrangian density
\begin{equation}
\mathcal{L}_m=D_\mu\phi(D^\mu\phi)^*-\frac{1}{4}F_{\mu\nu}F^{\mu\nu}-\frac{\sigma}{2}(\phi\phi^*-\eta^2)^2,
\end{equation}
where $D_\mu\phi=\nabla_\mu \phi-ieA_\mu\phi$ is the covariant derivative associated with the complex scalar field. The field strength
tensor is $F_{\mu\nu}=\nabla_\mu A_\nu -\nabla_\nu A_\mu=\partial_\mu A_\nu-\partial_\nu A_\mu$,
of the U(1) gauge potential $A_\mu$ with the coupling constant $e$. During the symmetry breaking process both the Higgs and the gauge fields acquire masses given by $M_{H}=\sqrt{2 \lambda} \eta$ and $M_{W}=\sqrt{2}e \eta$, respectively. The parameter $\sigma$ denotes the self-coupling of the scalar field, whereas $\eta$ is its vacuum expectation value. As it is known the corresponding energy-momentum tensor is obtained from the
Lagrangian through the relation below
\begin{equation}
T_{\mu\nu}=2\frac{\delta \mathcal{L}_m}{\delta g^{\mu\nu}}-g_{\mu\nu}\mathcal{L}_m.
\label{EM-definition}
\end{equation}
This leads to the explicit form of the energy-momentum tensor given by
\begin{eqnarray}
T_{\alpha\beta}&=&-g_{\alpha\beta}\left\{ (D_\mu\phi)(D^\mu\phi)^* -\frac{1}{4}F_{\mu\nu}F^{\mu\nu}-\frac{\sigma}{2}(\phi\phi^*-\eta^2)^2
\right\}\nonumber\\
&+&(D_\alpha \phi)(D_\beta\phi)^* +(D_\beta \phi)(D_\alpha \phi)^* - F_{\alpha\gamma}F_\beta^\gamma,
\end{eqnarray}
where the Higgs and gauge fields are \cite{nielsen}
\begin{eqnarray}
\phi(r,\varphi)=\eta f(r)e^{i\varphi} \ \ \ , \ \ \ A_\mu dx^\mu = \frac{1}{e}[1-P(r)]d\varphi.
\end{eqnarray}
The most general static cylindrically symmetric line element, invariant under boosts along the z-direction is 
\begin{equation}
\label{lineEl}
ds^2=N^2dt^2-dr^2-L^2d\varphi^2-N^2dz^2,
\end{equation}
where $N=N(r)$ and $L=L(r)$. 

We can verify that the symmetry of the system under consideration imposes to the four-vector $\lambda_{\mu}$ the particular form $(0,\lambda_r,0,0)$, with $\lambda_r=\textrm{const}$. 
The components  of the Ricci tensor associated with (\ref{lineEl}) are  
\begin{eqnarray}
\label{rmn}
R_t^t=\frac{(LNN')'}{N^2L} \ \ \ , \ \ \ R_r^r=\frac{2N''}{N}+\frac{L''}{L} \ \ \ , \ \ \ R_\varphi^\varphi=\frac{(N^2L')'}{N^2L}
\ \ \ , \ \ \ R_z^z=R_t^t.
\end{eqnarray}
Whilst for $K_{\mu\nu}$ we have
\begin{eqnarray}
\label{kmn2}
K_t^t=\lambda_r\frac{N'}{N}=K_z^z \ \ \ , \ \ \ K_r^r=\lambda_r\left(\frac{2N'}{N}+\frac{L'}{L}\right) \ \ \ , \ \ \
K_\varphi^\varphi=\lambda_r\frac{L'}{L}.
\end{eqnarray}

It is convenient to introduce the following redefinitions
\begin{eqnarray}
r\rightarrow \frac{r}{e \eta} \ \ \ , \ \ \ L\rightarrow \frac{L}{e\eta} \ \ \ , \ \ \ \lambda_r\rightarrow (e\eta) \lambda_r ,
\end{eqnarray}
with the radial distance $r$ being measured in units of $M_W/\sqrt{2}$. Besides, the Lagrangian density, $\mathcal{L}_m\rightarrow \mathcal{L}_m/(\eta^4e^2)$,
is written in terms of the dimensionless coupling constants below
\begin{eqnarray}
\alpha=8\pi G\eta^2=8\pi\frac{\eta^2}{M_{\rm pl}^2} \ \ \ , \ \ \ \beta=\frac{\sigma}{e^2}=\frac{M_H^2}{M_W^2},
\end{eqnarray}
where $M_{\rm pl}$ is the Planck mass. 

Using the metric (\ref{lineEl}) in (\ref{EM-definition}) we find the non-vanishing components of the energy-momentum tensor below
\begin{eqnarray}
\label{emt}
T_t^t&=&(f')^2+\frac{f^2P^2}{L^2}+\frac{(P')^2}{2L^2}+\frac{\beta}{2}\left(f^2-1\right)^2 \  \ , \  \ T_t^t=T_z^z,\\
\nonumber\\
T_r^r&=&-(f')^2+\frac{f^2P^2}{L^2}-\frac{(P')^2}{2L^2}+\frac{\beta}{2}\left(f^2-1\right)^2,\\
\nonumber\\
T_\varphi^\varphi&=&(f')^2-\frac{f^2P^2}{L^2}-\frac{(P')^2}{2L^2}+\frac{\beta}{2}\left(f^2-1\right)^2
\end{eqnarray}

while the trace $T\equiv T_\mu^\mu$ is
\begin{eqnarray}
T&=&2\left[(f')^2+\frac{f^2P^2}{L^2}+\beta(f^2-1)^2\right].
\end{eqnarray}
From $T_t^t$ and $T_z^z$ it is possible to obtain two important properties of the string which are its tension, $\tau$, and linear energy density. These two quantities are, respectively, given by
\begin{equation}
\label{tau} 
\tau=\int_{0}^{\infty} \int_{0}^{2 \pi} T_{z}^{z} \sqrt{g^{(2)}} d \rho d \varphi,
\end{equation}
and
\begin{equation}
\label{mu} 
\mu=\int_{0}^{\infty} \int_{0}^{2 \pi} T_{t}^{t} \sqrt{g^{(2)}} d \rho d \varphi.
\end{equation}
However, due to the boost invariance in the $z-$direction, we shall have $\mu=\tau$. One denotes deficit angle by $\delta$, which can be defined in terms of the metric function $L(r)$ as follows
\begin{equation}
\label{delta}
\delta =2\pi \left[1-L'(\infty)\right].
\end{equation}
\subsection{The field equations}

When we use (\ref{lineEl}), (\ref{rmn}) and (\ref{kmn2}), along with (\ref{emt}), in the modified field equations (\ref{feqs}) we are left with the following system of nonlinear differential equations
\begin{eqnarray}
\label{feqs1}
\frac{(LNN')'}{N^2L}+\lambda_r\frac{N'}{N}&=&\alpha\left[\frac{(P')^2}{2L^2}-\frac{\beta}{2}(f^2-1)^2\right],\\
\nonumber \\
\label{feqs2}
\frac{2N''}{N}+\frac{L''}{L}+\lambda_r\left(\frac{2N'}{N}+\frac{L'}{L}\right)&=&\alpha\left[-2(f')^2-\frac{(P')^2}{2L^2}
-\frac{\beta}{2}(f^2-1)^2\right],\\
\nonumber \\
\label{feqs3}
\frac{(N^2L')'}{N^2L}+\lambda_r\frac{L'}{L}&=&\alpha\left[-\frac{2f^2P^2}{L^2}-\frac{(P')^2}{2L^2}
-\frac{\beta}{2}(f^2-1)^2\right].
\end{eqnarray}
The variables involved in the problem are subject to the following boundary conditions
\begin{eqnarray}
\label{bc1}
L(0)=0,\;\;\;L'(0)=1,\;\;\;N(0)=1,\;\;\;N'(0)=0,\;\;\;f(0)=0,\;\;\;P(0)=1,
\end{eqnarray}
that guarantee a regular solution at the $z-$axis. Besides, we admit that the matter fields reach their respective vacuum expectation values far away from the string, which means
\begin{equation}
\label{bc2} 
f(r\rightarrow \infty)\rightarrow 1,\;\;P(r\rightarrow \infty)\rightarrow 0. 
\end{equation}

We can combine properly (\ref{feqs1})-(\ref{feqs3}) so that a new constraint is obtained:
\begin{eqnarray}
\frac{2N'L'}{NL}+\frac{(N')^2}{N^2}&=&\alpha\left[(f')^2+\frac{(P')^2}{2L^2}-\frac{f^2P^2}{L^2}
-\frac{\beta}{2}(f^2-1)^2\right],
\label{rrcontrib}
\end{eqnarray}
which is going to be useful to us later on. Since the correction $\lambda_{\mu}s^\mu$ in (\ref{ADL}) does not depend on the Higgs and gauge fields, this term is unaffected by the extremisation of the action with respect to these fields, so that the dynamical equations for the both are strictly the same as the GR ones, namely  
\begin{eqnarray}
\label{eqF}
f''+\frac{L'}{L}f'+\frac{2N'}{N}f'-\frac{fP^2}{L^2}-\beta f(f^2-1)=0
\end{eqnarray}
and 
\begin{eqnarray}
\label{eqP}
\frac{P''}{L^2}+\frac{2N'}{N}\frac{P'}{L^2}-\frac{L'P'}{L^3}-\frac{2Pf^2}{L^2}=0.
\end{eqnarray}
On the other hand, the modified conservation law (\ref{nclaw}) gives
\begin{eqnarray}
\label{nclaw1}
2\lambda_r\left[\frac{2N'L'}{NL}+\frac{(N')^2}{N^2}\right]&=&\alpha \left[-2f' \left(f''+\frac{L'}{L}f'+\frac{2N'}{N}f'
-\frac{fP^2}{L^2}-\beta f(f^2-1)\right)\right.\nonumber\\
&-&\left. P'\left(\frac{2N'}{N}\frac{P'}{L^2}+\frac{P''}{L^2}-\frac{L'P'}{L^3}-\frac{2Pf^2}{L^2}\right)\right].
\end{eqnarray}
When used in the equation above, (\ref{eqF}) and (\ref{eqP}) lead straightforwardly to two possible conditions for the metric functions $N(r)$ and $L(r)$, for $\lambda_r\neq 0$:
\begin{equation}
\label{n1}
N(r)=\textrm{const.}
\end{equation}
or
\begin{equation}
\label{l1}
\frac{2L'}{L}=-\frac{N'}{N}.
\end{equation}
For the sake of simplicity we are going to stick to the first case, in which the metric function $N(r)$ can be merely written as $N(r)=1$. It is easy to see that such a choice implies in the so-called BPS limit for which $\beta=1$ \cite{Bogomol'nyi,PS:1975}. Additionally, $N(r)=1$ makes much simpler the set of equations of motion for the Abelian-Higgs string which now becomes
\begin{eqnarray}
\label{bps}
f'=\frac{P}{L}f,\;\;\;P'=L(f^2-1)\;\;\;\textrm{and}\;\;\;L''+\lambda_rL'=-\alpha L\left[(f^2-1)^2+\frac{2P^2f^2}{L^2}\right].
\end{eqnarray}
It is worth comparing the present study with the results presented in \cite{betti}. There the authors investigate Abelian-Higgs strings within another non-conservative framework, namely the Rastall theory \cite{rastall} \footnote{Contrary to the theory under analysis, Rastall gravity is a theory where a violation of the traditional conservation law is assumed in a purely phenomenological way, without a proper variational formulation.}. The main difference we may remark between the both studies is related to the gauge-Higgs couplings ratio, namely the constant $\beta$. In our case, as it is shown above, the consistency imposed by the dynamical equations sets $\beta=1$, which means that the gauge-Higgs couplings (likewise the gauge-Higgs masses) are necessarily equal. On the other hand, in Rastall case the $\beta$ parameter is strongly dependent of the Rastall parameter, thus it can assume a much wider range of values. Since it has direct consequences on the physical properties of both the gauge and Higgs fields, it means that the structure of the fields may be heavily affected by the type of non-conservative gravity in which they are studied.   
\section{Numerical results}
 
\begin{figure}[h]
  \centering
  \subfloat[]{\includegraphics[width=0.5\textwidth]{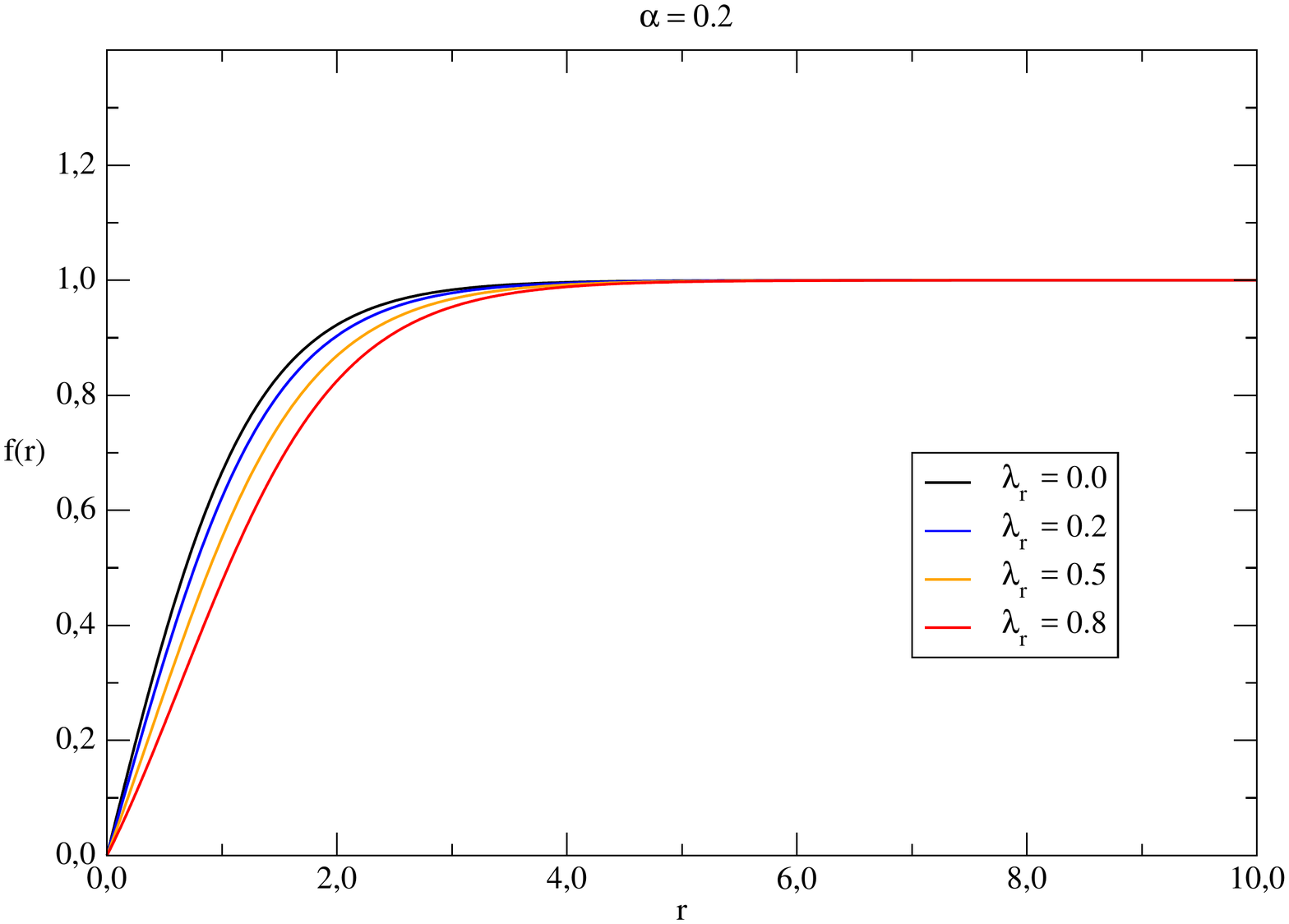}\label{fig1}}
  \hfill
  \subfloat[]{\includegraphics[width=0.5\textwidth]{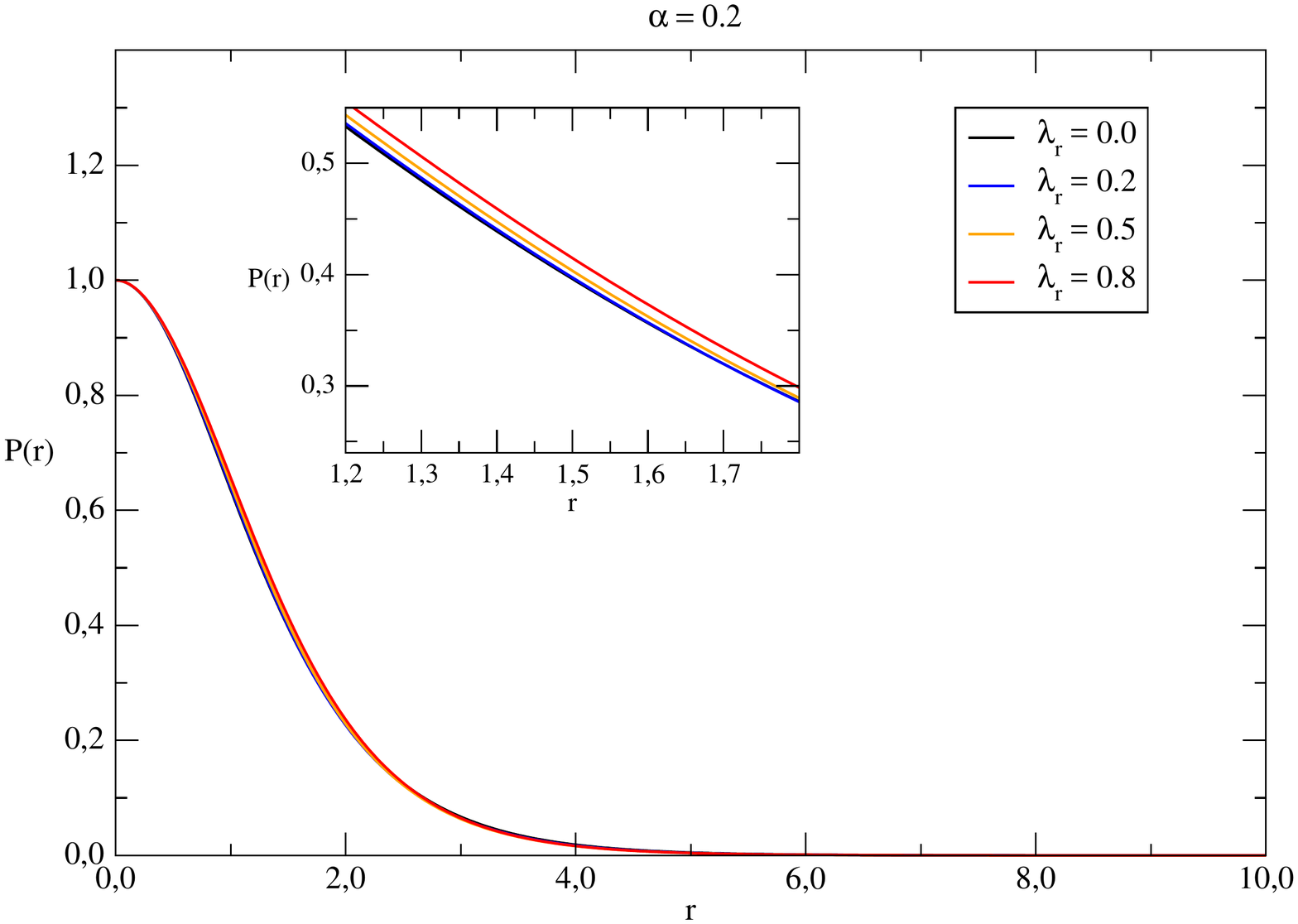}\label{fig2}}
  \caption{
  In the left panel, we present the behavior for the Higgs
  field as function of $r$. In the right panel, the gauge field, $P(r)$, is shown. For the both situations we fix $\alpha=0.2$ and take successively the values $\lambda_r = 0.0, 0.2, 0.5, 0.8$}
\end{figure}

\begin{figure}[h]
	\centering
	\subfloat[]{\includegraphics[width=0.5\textwidth]{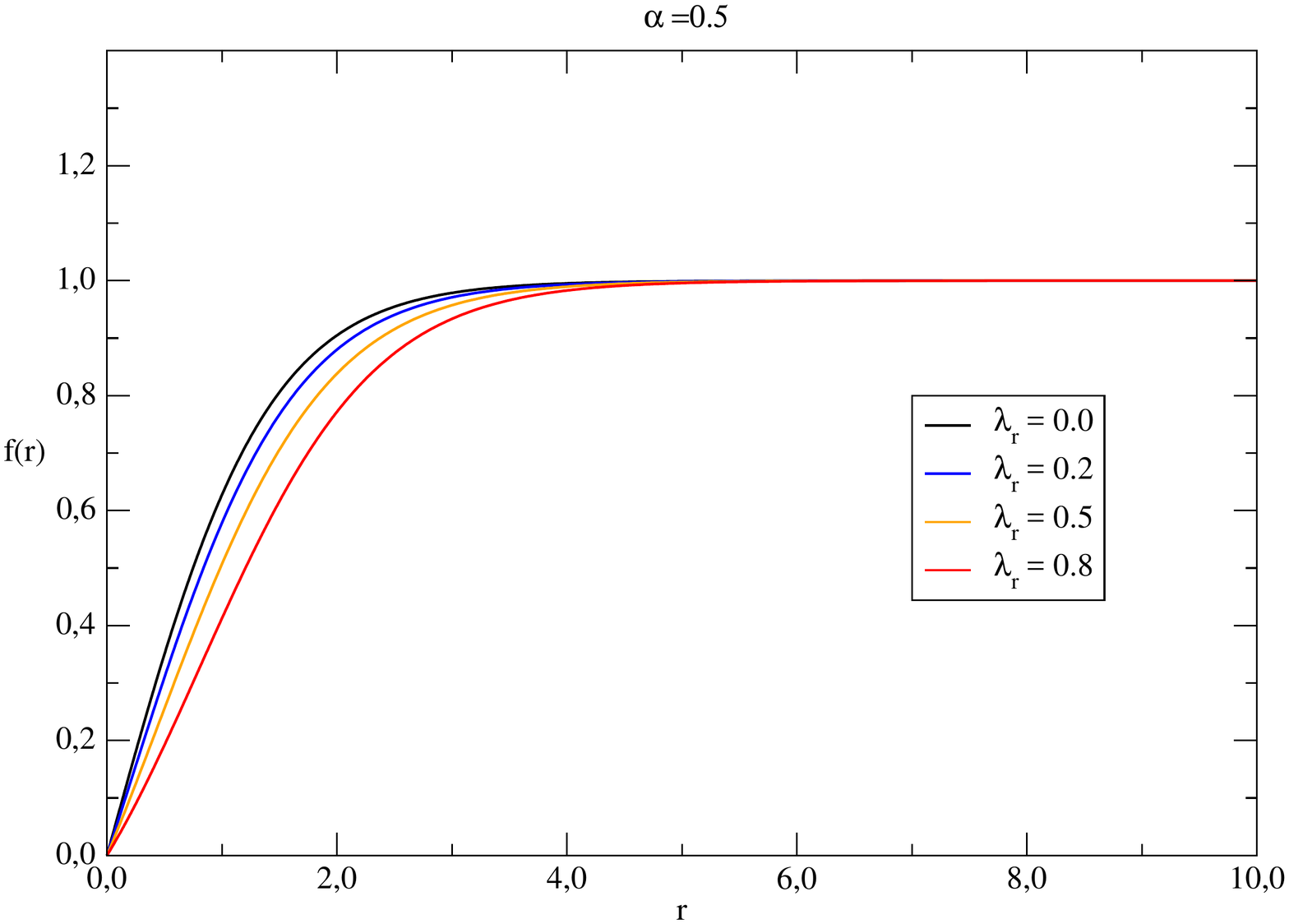}\label{fig3}}
	\hfill
	\subfloat[]{\includegraphics[width=0.5\textwidth]{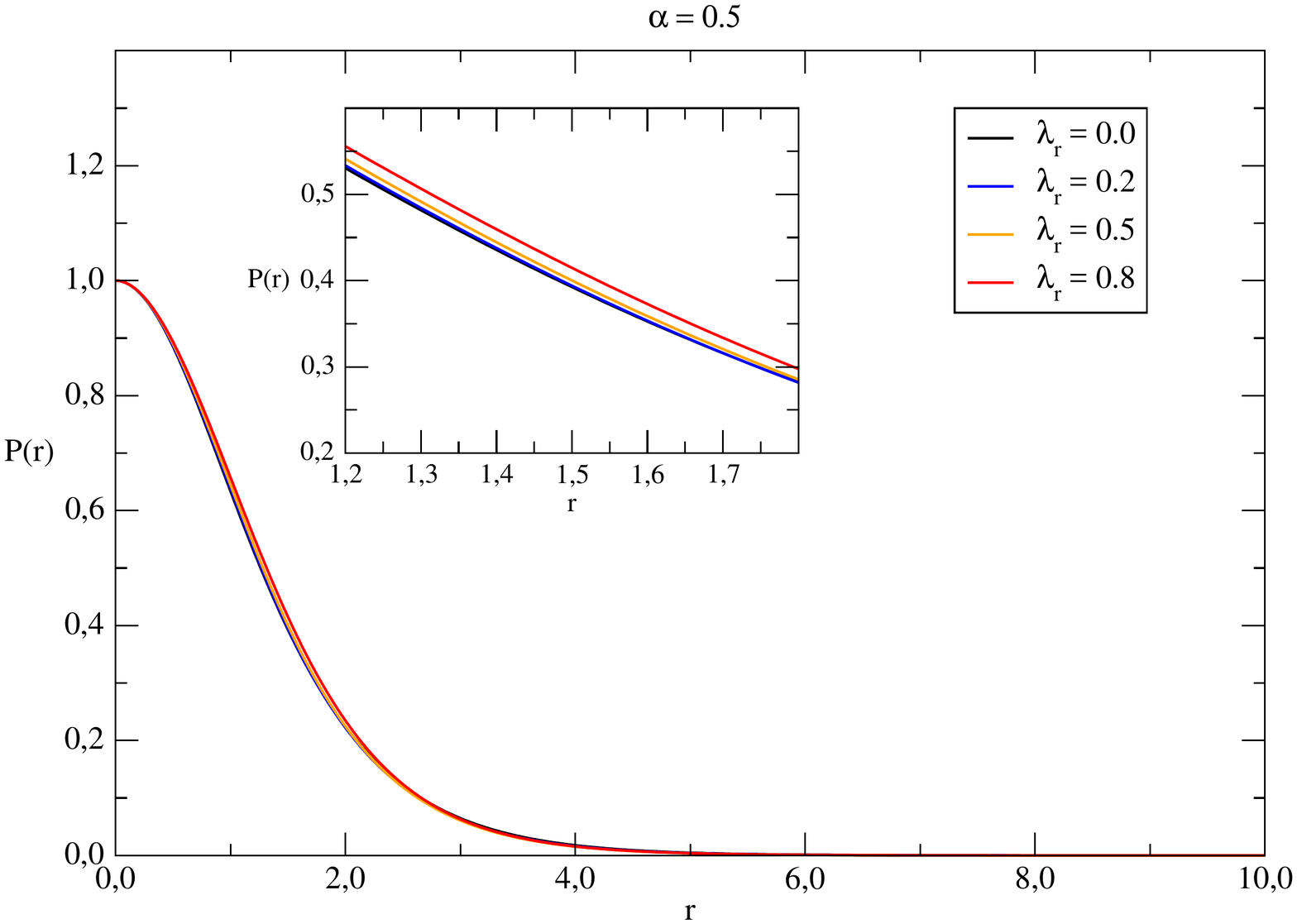}\label{fig4}}
	\caption{In the left panel, we present the behavior for the Higgs
		field as function of $r$ considering the parameters $\alpha  = 0.5$ and $\lambda_r = 0.0, 0.2, 0.5, 0.8$. In the right panel we exhibit the gauge field profile for the same $\alpha$'s and $\lambda_r$'s values.}
\end{figure}
\begin{figure}[h]
	\centering
	\subfloat[]{\includegraphics[width=0.5\textwidth]{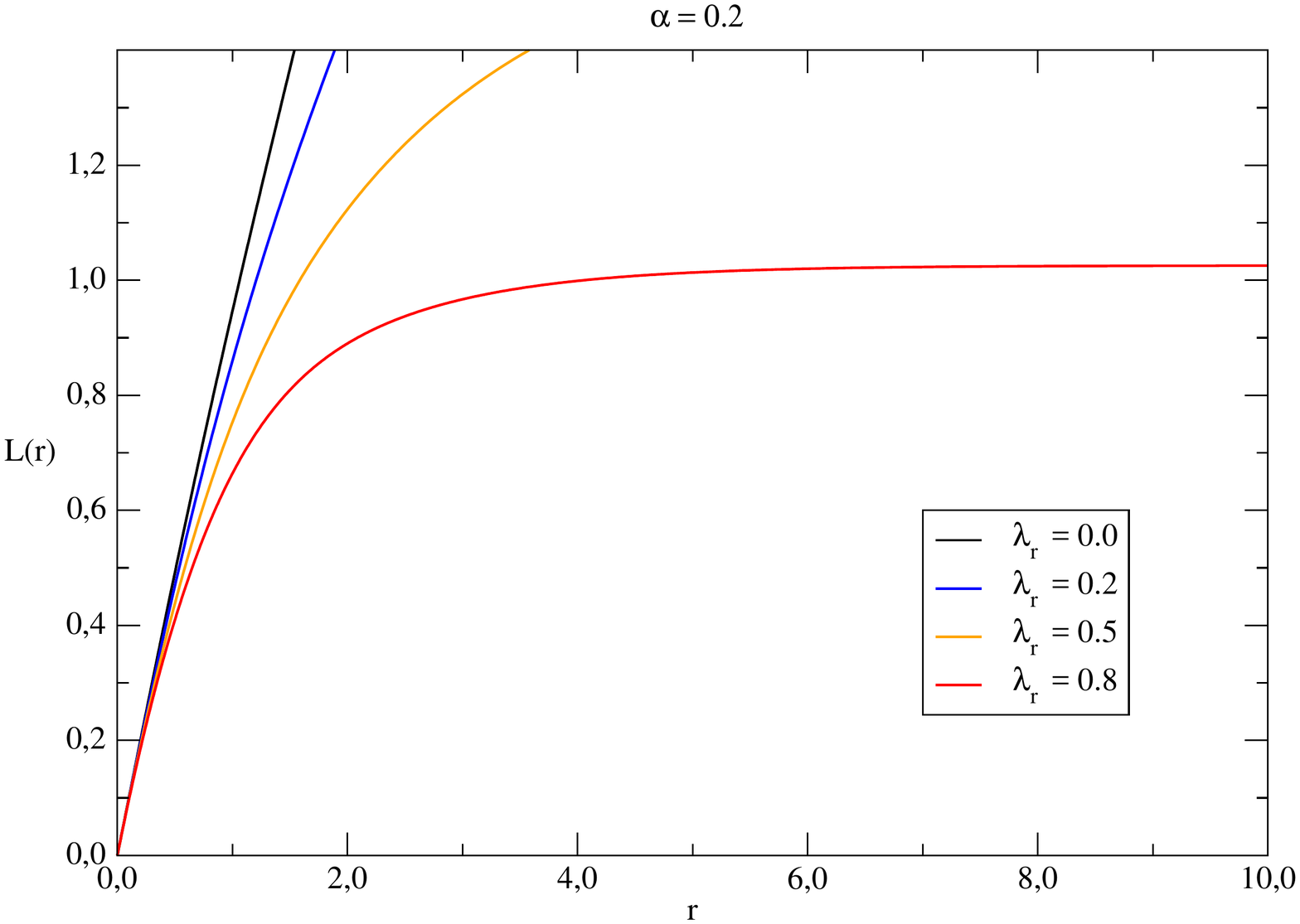}\label{fig5}}
	\hfill
	\subfloat[]{\includegraphics[width=0.5\textwidth]{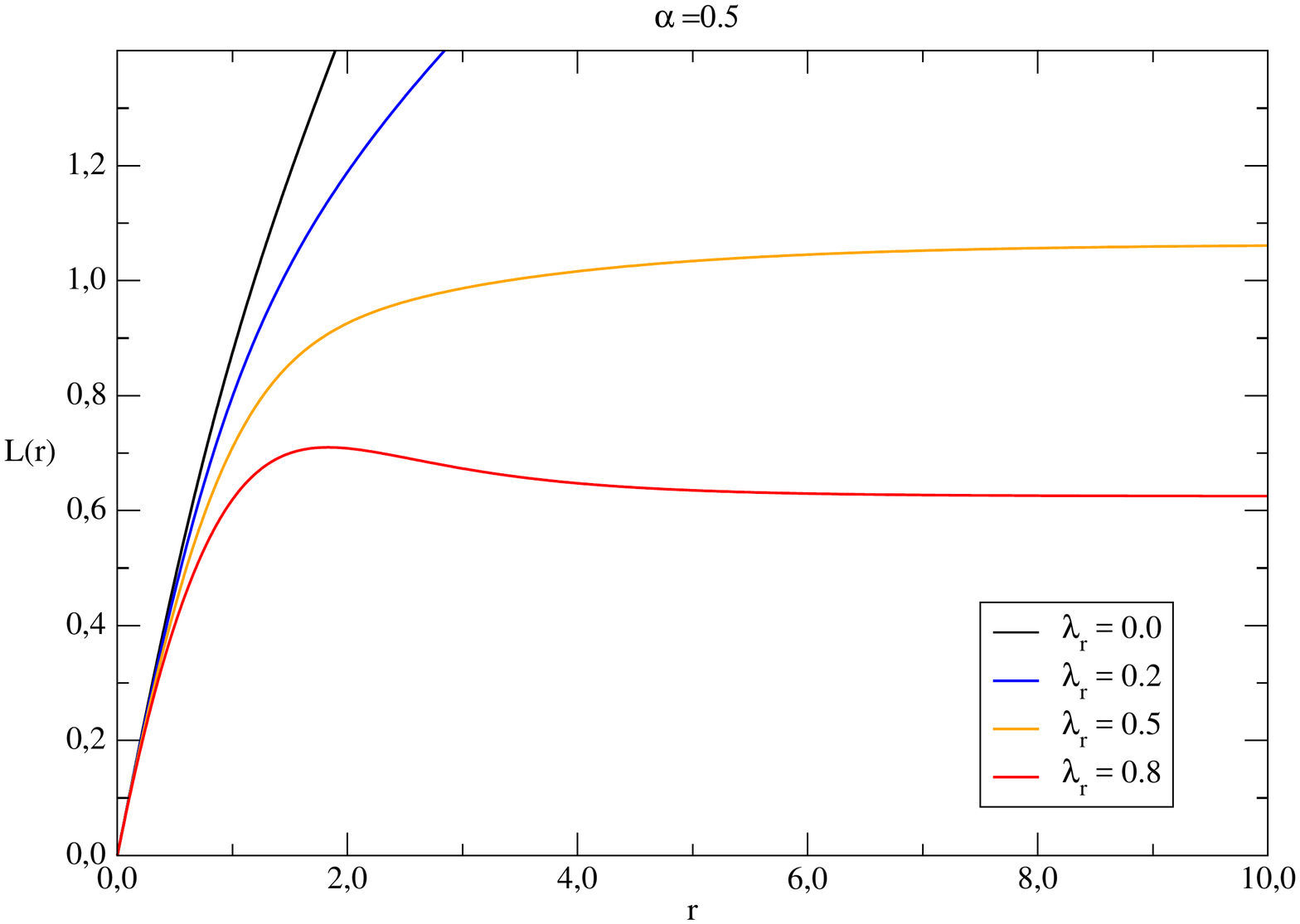}\label{fig6}}
	\caption{The comparison between the behavior of the metric function $L$ as functions of $r$. In the left panel considering $\alpha = 0.2$ and the right panel considering $\alpha = 0.5$. In the both plots we cover the cases $\lambda_r = 0.0, 0.2, 0.5, 0.8$. 
		}
\end{figure}

\begin{figure}[h]
	\centering
	\subfloat[]{\includegraphics[width=0.5\textwidth]{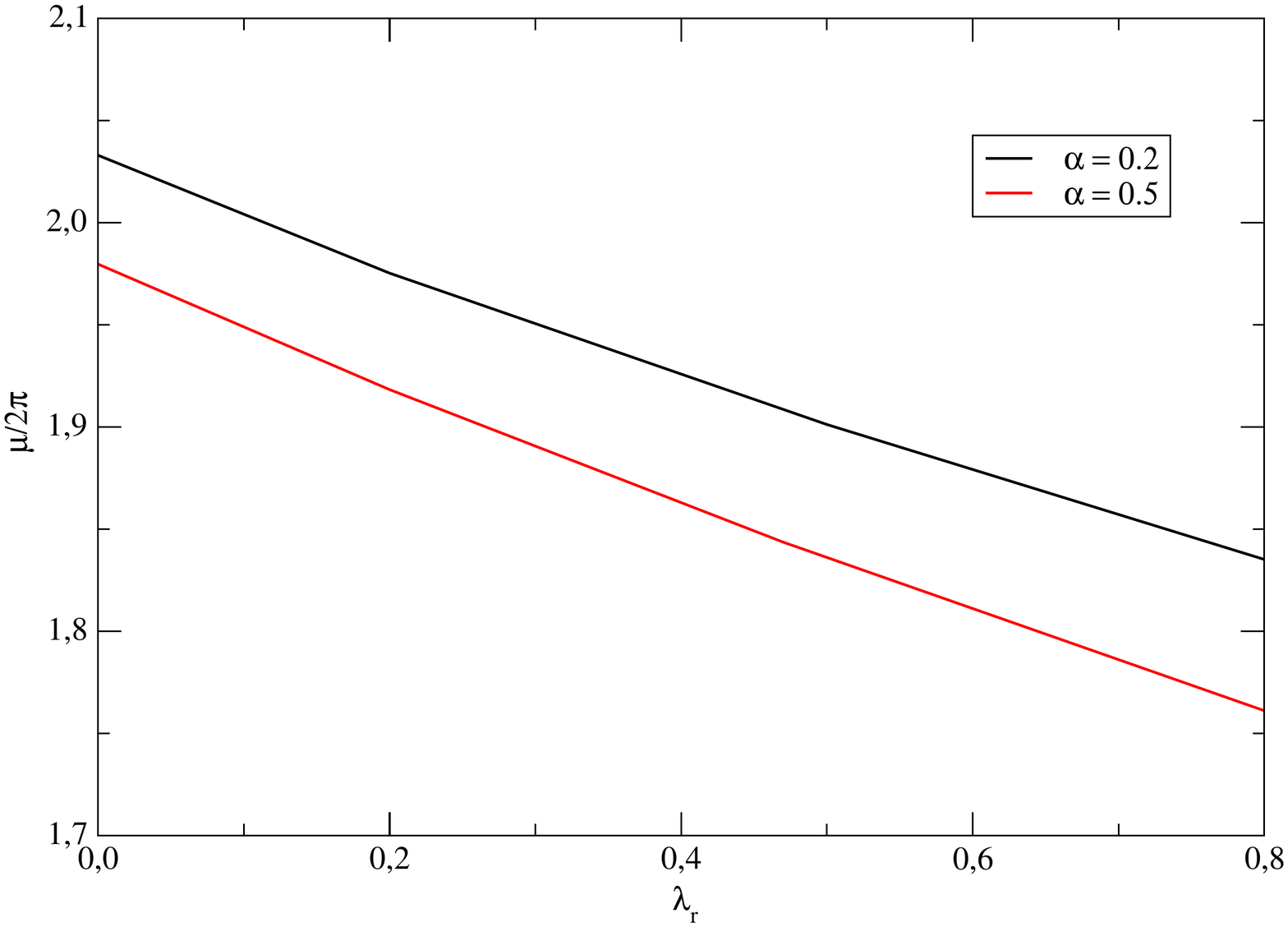}\label{fig7}}
	\hfill
	\subfloat[]{\includegraphics[width=0.5\textwidth]{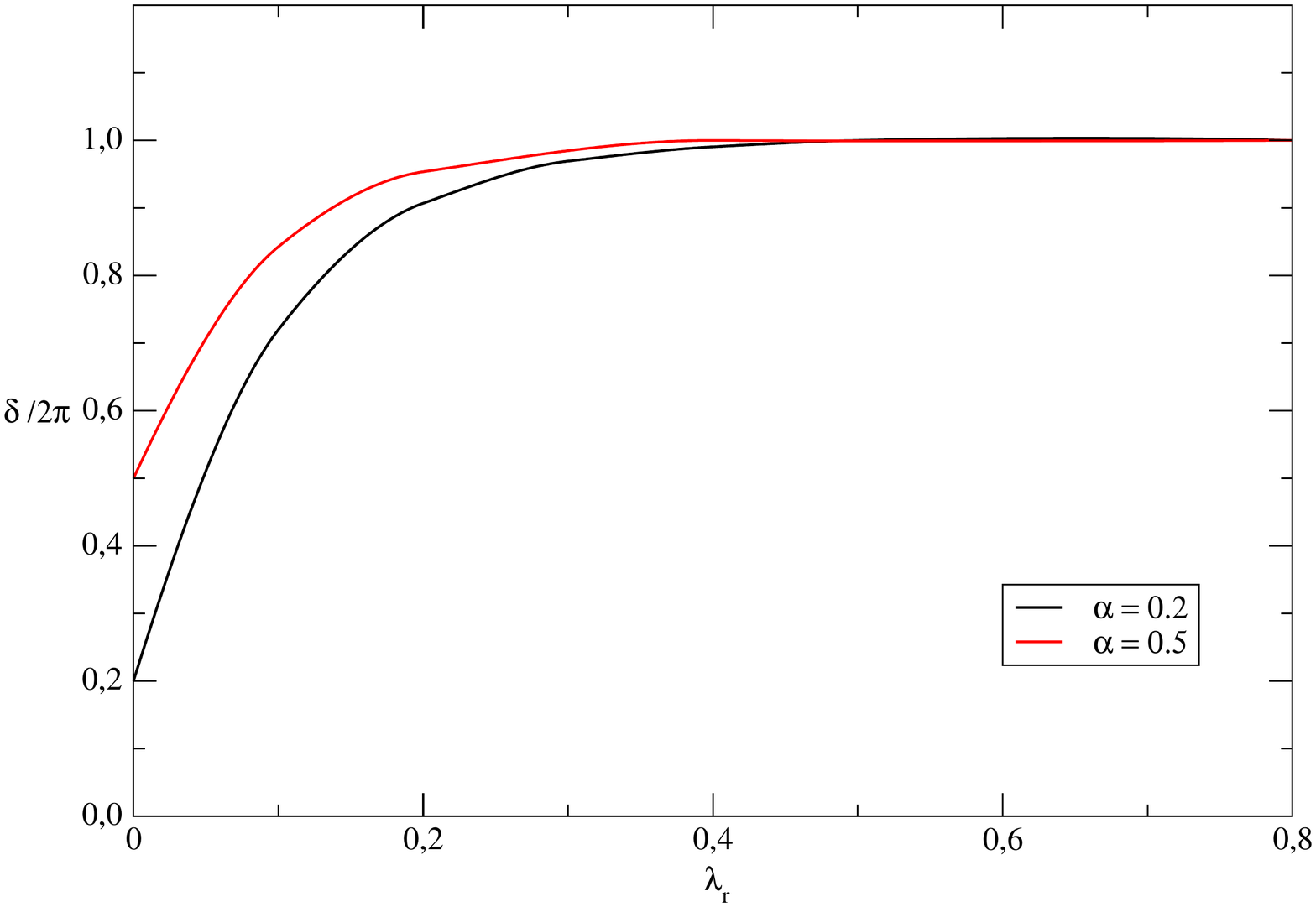}\label{fig8}}
	\caption{In the left panel, we present the behaviour of density of the energy per unit of $2\pi$ as a function of $\lambda_r$. In the right panel,  the behaviour of the planar deficit angle in units of $2\pi$ as a function of $\lambda_r$.}
\end{figure}
Now we are going to describe the numerical results obtained in this work. We have integrated the system of nonlinear differential equations (\ref{bps}) considering the set of boundary conditions (\ref{bc1}) and (\ref{bc2}). This was performed by means of the ODE solver COLSYS \cite{colsys}. Relative erros of the solutions are usually comprised within the order of $10^{-8}$ to $10^{-10}$ (sometimes even better).

In the Fig. 1 we present the plots of the Higgs and gauge fields, $f(r)$ and $P(r)$, for different values of $\lambda_r$, fixing $\alpha=0.2$. In the Fig. 2 we report a similar analysis, however for $\alpha=0.5$. Notice that increasing the value of $\lambda_r$ makes the both fields $f(r)$ and $P(r)$ get more and more spread around the string axis, causing an effective increasing of the string width. However, these fields revealed almost insensitive to the $\alpha$ parameter. Additionally, we observe a common tendency of both fields to reach ``slowly" its vacuum expectation value for higher values of $\lambda_r$. 

In the Fig. 3 we present the behaviour of metric function $L(r)$ considering again the parameters $\alpha =0.2$ and $\alpha=0.5$. For the both cases, we consider $\lambda_r = 0.0, 0.2, 0.5$ and $0.8$. We notice that $L(r)$ is less sensitive to the $\lambda_r$'s effects at small distances, although a tiny discrepancy is perceiveable for differents $\lambda_r$. However, the deviation from GR shows up indeed more pronounced for large $r$ , for which $L(r)$ has its slope decreased when one takes higher and higher values for the parameter $\lambda_r$. This behaviour indicates that stronger deviations from GR shall lead to larger values for the deficit angle.         

The behaviours of the linear energy density (or mass per unit length of the string) and planar angle deficit are shown in the Fig. 4. In the left panel we plot the linear energy density as a function of $\lambda_r$ considering the cases $\alpha =0.2$ and $\alpha =0.5$. The right panel displays the planar angle deficit as a function of $\lambda_r$, for these same two values of $\alpha$. We notice a decreasing of the linear energy density as the $\lambda_r$ parameter, which can be seen as a direct consequence of the dissipative effect one expects to observe within such an extended gravity. On the other hand, for the both cases of $\alpha$ the planar angle deficit increases as $\lambda_r$ takes higher values. We also see that the influence of $\alpha$ weakens as $\lambda_r$ is enhanced. Additionally, we notice the existence of a critical value for $\lambda_r$ about $\lambda_r^{cr}\approx 0.5$, at which the deficit angle reaches its maximum value, $2 \pi$. 
From this $\lambda_r^{cr}$ onwards, the deficit angle saturates at its maximum value, what spoils any suitable description of a Abelian-Higgs string within such a gravitational theory for $\lambda_r\gtrsim \lambda_r^{cr}$. Therefore, this modified gravity is only able to accommodate a proper description of a gravitating Abelian-Higgs string if the departure of GR is not so large. 

\section{Analytical treatment}
After a full numerical treatment of the system (\ref{bps}) we present in this section the main analytical findings of our study.
\subsection{Solution close to the string axis}
Using the set of boundary conditions (\ref{bc1}) we can write the approximate form for the system of equations (\ref{bps}) which will by turn provide solutions for $L(r)$, $P(r)$ and $f(r)$ in the regions very close to the string axis ($r\sim 0$). In this regime, due to the condition $f(0)=0$, it is reasonable to keep only linear terms in $f(r)$ and neglect all the possible higher contributions of these function. This implies in a approximative form for the system (\ref{bps}) given by: 
\begin{eqnarray}
\label{bps2}
f'=\frac{P}{L}f,\;\;\;P'\simeq -L\;\;\;\textrm{and}\;\;\;L''+\lambda_rL'\simeq-\alpha L,
\end{eqnarray}
which gives the following set of solutions
\begin{equation}
\label{L1}
L(r)\backsimeq r-\frac{\lambda_r}{2}r^2+O(r^3),
\end{equation}
\begin{equation}
\label{P1}
P(r)\backsimeq 1-\frac{r^2}{2}+\frac{\lambda_r}{6}r^3+O(r^4),
\end{equation}
and
\begin{equation}
\label{f1}
f(r)\backsimeq r+\frac{\lambda_r}{2} r^2+O(r^3).
\end{equation}
As expected, the behaviours both of the matter fields and the metric function showed above are clearly endorsed by the full numerical analysis we have performed in the previous section. As the plots (\ref{fig1}), (\ref{fig3}), (\ref{fig5}) and (\ref{fig6}) show us, the profiles of $f(r)$ and $L(r)$ are only considerably affected by $\lambda_r$ as one departs the string's vicinity, with a significant influence only from the $r^2$ powers onwards. As such, the closer to the string axis the more these fields look like their GR counterparts, with $f(r)$ being a bit more sensitive than $L(r)$ to the modification of gravity in this close-to-string regime. On the other hand, we observe that $\lambda_r$'s impact on $P(r)$ behaviour is even weaker, with a dependence on this parameter becoming evident just from the $r^3$ power onwards. In fact, one notices from Figs. (\ref{fig2}) and (\ref{fig4}) that this tiny dependence of $P(r)$ upon $\lambda_r$ is also verified even in the full regime.     

\subsection{Vacuum solution}
One assumes the cosmic string is a localized configuration whose corresponding matter fields shall vanish far away its core and the spacetime of the string will reduce to the cylindrically symmetric vacuum solution of the respective gravitational theory. In GR the vacuum for this symmetry is described by the Kasner solutions \cite{Kasner}, however here one expects a new set of metric functions modified by the contribution of the parameter $\lambda_r$. In this vein, we integrated the field equations (\ref{feqs1})-(\ref{feqs3}) in empty space and obtained an exact result given by 
\begin{equation}
\label{Nr}
N(r)=N_1\left(\frac{1-e^{-\lambda_r r}}{\lambda_r}\right)^{a}
\end{equation}
and
\begin{equation}
\label{Lr}
L(r)=L_1\left(\frac{1-e^{-\lambda_r r}}{\lambda_r}\right)^{b},
\end{equation}
where $N_1$ and $L_1$ are constants. For the sake of simplicity, we shall restrict our attention to regular solutions whose behaviour is well-defined in all the space. This leads us to assume $\lambda_r$'s sign as positive, since a negative one would imply in a naked singularity at infinity, as one can clearly see from (\ref{Nr}) and (\ref{Lr}). Notice that in the limit when $\lambda_r \rightarrow 0$, the solution above recover the Kasner one $N(r)\sim N_1 r^{a}$ and $L(r)\sim L_1 r^{b}$. The relation between the exponents $a$ and $b$ is 
\begin{equation}
\label{kas}
2a^2+b^2=2a+b=1,
\end{equation}
which is found by means of the vacuum equations arising from (\ref{feqs1}). This relation is the same as the one found in the GR context \cite{Kasner} and leads to the two possible solutions below
\begin{equation}
\label{kas1}
(a,b)=(0,1)\;\;\;\textrm{and}\;\;\;(a,b)=\left(\frac{2}{3},-\frac{1}{3}\right).
\end{equation}
In fact the first case is particularly interesting for the study of cosmic strings as it contains solutions with deficit angle. So, we shall focuse on this case, in which $(a,b)=(0,1)$. Notice that when setting $b=1$ in (\ref{Lr}) one finds $L'(\infty)=0$, leading to $\delta=2\pi$, which is the critical limit acceptable for the deficit angle. This indicates that for $r$ large the surface $(t,z)$ = const of the vacuum geometry approaches asymptotically a cusp. 

\subsection{The string as an extended source}
It is common to consider the string as a line source, meaning that all the matter distribution characterising the defect is concentrated at the central axis. This leads one to model the cosmic string in terms of a delta-like energy-momentum tensor, which arises as the suitable way to represent a conical singularity lying at the origin, with a locally flat spacetime being achieved in the outside region. However, such hypothesis is somewhat unrealistic, as it would bring an undesirable divergence to the model which possibly would cause difficulties to the practical treatment of this problem. A simple and direct attempt to circumvent this inconvenient feature could be assuming that the defect is enclosed in a finite radius, $\rho_0$, inside of which its energy density is distributed. Let us consider that the inner region is described by the following metric tensor
\begin{equation}
\label{inner} 
ds^2=A^2dT^2-d\rho^2-B^2d\varphi^2-A^2dZ^2,
\end{equation}
where $A=A(\rho)$ and $B=B(\rho)$. The energy distribution of the string is parametrized by a radial function $\epsilon(\rho)$. So, this energy-momentum tensor could be written as
\begin{equation}
\label{tem1}
{\cal T}_{\mu}^{\nu}=\epsilon(\rho) \textrm{diag}(1,0,0,1).
\end{equation}
For this configuration the modified field equations become 
\begin{eqnarray}
\label{FEqs}
\frac{(BAA')'}{A^2B}+\lambda_r\frac{A'}{A}&=&0,\\
\nonumber \\
\frac{2A''}{A}+\frac{B''}{B}+\lambda_r\left(\frac{2A'}{A}+\frac{B'}{B}\right)&=&-8 \pi G \epsilon(\rho),\\
\nonumber \\
\frac{(A^2B')'}{A^2B}+\lambda_r\frac{B'}{B}&=&- 8 \pi G \epsilon(\rho),
\end{eqnarray}
where the prime here is denoting derivative with respect to $\rho$. It is easy to verify that the system above also admits the same two family of solutions (\ref{n1}) and (\ref{l1}). From the both cases we shall focuse on that one which is simpler and also more interesting for the study of cosmic strings, namely $A(\rho)=\textrm{const.}$ This case makes the remaining field equations to be reduced into a single one given by
\begin{equation}
\label{Br}
B''(\rho)+\lambda_r B'(\rho) +8\pi G \epsilon(\rho)B(\rho)=0.
\end{equation}
To solve (\ref{Br}) we should in principle specify $\epsilon(\rho)$. However, in order to obtain a closed solution for this equation and capture the immediate effect of the parameter $\lambda_r$, we can simplify the problem by assuming a model where the energy density is almost uniform, $\epsilon (\rho)\sim \epsilon_0$. This implies in the following solution \footnote{Although the results obtained in this section so far are valid for any value of $\lambda_r$, from now on we shall concentrate in a specif case where the condition $\frac{\lambda_r^2}{4}< 8 \pi G \epsilon_0$ is assumed, implying in (\ref{solLr}).}
\begin{equation}
\label{solLr} 
B(\rho)=e^{-\lambda_r \rho/2}\left[a_0\; \textrm{cos}(\rho/\rho_{*})+b_0\; \textrm{sin}(\rho/\rho_{*})\right],
\end{equation}
where $a_0$ and $b_0$ are integration constants and $\rho_{*}\equiv\left[8\pi G \epsilon_0 - \frac{\lambda_r^2}{4}\right]^{-1/2}$. Notice that for $\lambda_r=0$ we recover the Hiscock's solution \cite{Hisc}, obtained in the context of GR. This solution was also derived independently by J. R. Gott III in \cite{gott}, where the author discuss possible consequences of this solution on the gravitational lens effect. 

Let us recall that (\ref{solLr}) is subject to the boundary condition $B(0)=0$ which imposes the fixing $a_0=0$. Besides, the constant $b_0$ can be determined by following the same Hiscock's claim: avoidance of conical singularity. This is achieved by setting $b_0=\rho_{*}$, so that the solution becomes  
\begin{equation}
\label{solLr1}
B(\rho)=e^{-\lambda_r \rho/2}\rho_{*}\; \textrm{sin}(\rho/\rho_{*}).
\end{equation}
The exterior metric is given by the vacuum solution (\ref{Lr}) [with $b=1$]:
\begin{equation}
\label{Lr1}
L(r)=L_1\left(\frac{1-e^{-\lambda_r r}}{\lambda_r}\right).
\end{equation}
We are looking for an exact solution holding in the entire space, so we must require that both the inner and the exterior solutions join together along the surface of the string. This means to assume continuity for $B(\rho)$ (interior) and $L(r)$ (exterior) and their respective first derivatives at $\rho=\rho_0$ and $r=r_0$. These two conditions provide, respectively, the equations below
\begin{equation}
\label{mc1}
L_1=\frac{e^{-\lambda_r \rho_0/2}\lambda_r}{2(1-e^{-\lambda_r r_0})} \rho_{*}\textrm{sin}(\rho_0/\rho_{*})
\end{equation}
and
\begin{equation}
\label{mc2}
L_1 e^{-\lambda_r r_0}=e^{-\lambda_r \rho_0/2}\left[-\frac{\lambda_r \rho_{*}}{2}\textrm{sin}(\rho_0/\rho_{*})+\textrm{cos}(\rho_0/\rho_{*})\right].
\end{equation}
In fact we are interested in determining the constant $L_1$ which carries the information about the deficit angle. This can be achieved from (\ref{mc1}) and (\ref{mc2}), by eliminating $r_0$ and reducing the both equations into a single which gives $L_1$ in terms of the remaining parameters of the model 
\begin{equation}
\label{mc3}
L_1 =e^{-\lambda_r \rho_0/2}\left[\frac{\lambda_r \rho_{*}}{2}\textrm{sin}(\rho_0/\rho_{*})+\textrm{cos}(\rho_0/\rho_{*})\right].
\end{equation}
Now we are able to assess the impact of the parameter $\lambda_r$ on the mass per unit lentgh (or linear energy density) of the string, which in terms of (\ref{tem1}) can be given by
\begin{equation}
\label{mu1} 
\tilde{\mu}=\int_{0}^{\rho_0} \int_{0}^{2 \pi} {\cal T}_{0}^{0} \sqrt{g^{(2)}} d \rho d \varphi,
\end{equation}
where $g^{(2)}_{ij}$ denotes the metric of the surface $(t,z)=$const. whose determinant is $B^2(\rho)$. This quantity plays a crucial role within the physics of cosmic strings, since it has connection with the energy scale of the symmetry breaking that produced the cosmic string. Given the energy scale at which the symmetry was broken, $\eta$, it is shown that $\mu \sim \eta^2$ \cite{VS}. This parameter is usually expressed in terms of a dimensionless quantity $G {\tilde{\mu}}^2$ which is indeed the main observable associated with the cosmic strings. From (\ref{mu1}) one finds
\begin{equation}
\label{mu2}
4 G \tilde{\mu}=1-e^{-\lambda_r \rho_0/2}\left[\frac{\lambda_r \rho_{*}}{2}\textrm{sin}(\rho_0/\rho_{*})+\textrm{cos}(\rho_0/\rho_{*})\right].
\end{equation}
which from (\ref{mc3}) leads to
\begin{equation}
\label{mu3}
4 G \tilde{\mu}=1-L_1.
\end{equation}
Since $\lambda_r$ is positive the equations above shows that the effective mass per unit length within this modified theory is smaller than its GR counterpart   
\begin{equation}
\label{muGR} 
\tilde{\mu}<\mu_{\tiny{GR}},
\end{equation}
where $4G\mu_{\tiny{GR}}=1-\textrm{cos}(\rho_0/\tilde{\rho})$ and $\tilde{\rho}\equiv\left(8\pi G \epsilon_0\right)^{-1/2}$. So, the geometric dissipative effects coming from such a deviation from GR contributes to a decreasing of the linear energy density of the cosmic string. As the main observable associated with cosmic string configurations, the mass per unit length may carry some information about the underlying gravitational theory. In GR, the gravitational lensing for cosmic strings reveals that the bending angle is proportional to $\mu$, as it is shown in \cite{vilenkin1}. So, the result (\ref{muGR}) makes the lensing effect a possible tool to distinguish the non-conservative gravity from the einstenian theory. Moreover, with aid of (\ref{mu3}) it is possible to write the exterior solution in terms of $\tilde{\mu}$ as follows
\begin{equation}
\label{extS}
ds^2=dt^2-dr^2-(1-4 G \tilde{\mu})^2\left(\frac{1-e^{-\lambda_r r}}{\lambda_r}\right)^{2}d\varphi^2-dz^2,
\end{equation}
which generalizes the exact solution originally obtained in the context of GR \cite{Hisc,gott} for arbitrary values of $\lambda_r$.   

\section{Conclusions and Perspectives}
In this work we have investigated some immediate consequences of a specific modified gravity on some cosmic string configurations. This alternative theory is characterized by the presence of dissipative effects emerging from first principles through a correction on the usual Einstein-Hilbert action, which makes the energy-momentum tensor respect a non-standard conservation law. We have obtained the respective dynamical equations for the Abelian-Higgs string and have found out that the modified conservation law of the energy-momentum tensor gives rise to a constraint which leads authomatically to $\beta\equiv\frac{M_H^2}{M_W^2}=1$, simplifying enormously the set of dynamical equations. This case can be seen as the corresponding BPS regime achievable within this model of gravity. Our following step was to integrate numerically these resulting system of nonlinear equations both for the metric and the matter fieds, analyzing the choices $\alpha=0.2$ and $\alpha=0.5$. We notice that the impact of $\alpha$ on $f(r)$ and $P(r)$ is quite small, contrarily to what occurs with the $\lambda_r$ parameter that clearly modifies the manner these fields get distributed around the string.
In particular, we verified that both $f(r)$ and $P(r)$ contributes to an increasing of the string's width as the modified gravity gets stronger and stronger. Besides, the consequences on the defict angle and the mass per unit length of the string are also adressed. From this analysis we found an upper bound for the $\lambda_r$ parameter given by $\lambda_r^{cr} \approx 0.5$ for which the deficit angle reaches its maximum value, $\delta_{max}=2 \pi$, leading to an undesirable cosmic string setup from the physical point of view. In our analytical approach we computed the solution close to the string which gave expected behaviours when compared with the numerical results. Moreover, we have obtained the exact solution in the absence of the matter fields, generalizing the standard Kasner type solution for the vacuum of cylindrical geometry of Einsteinian gravity. This solution was used in the subsection V.C, when we studied a gravitating cosmic string endowed with a finite radius. We computed the inner metric tensor and match it at the boundary with the vacuum solution outside. We have found a result generalizing the usual Gott-Hiscock solution for the non-conservative gravity. In special, we verified that the influence of $\lambda_r$ parameter in such a thick string configuration manifests as a decreasing in linear energy density of the string, what can be interpreted as a direct consequence of the dissipative effects emerging from this new theory of gravity leaving its signature on the main observable of the cosmic string physics. This work also leaves open issues that may be properly explored in a future opportunity, as for instance a more careful analysis of the class of solutions given by (\ref{l1}), which would also involve the use of numerical methods in a similar way to what we have done here. Furthermore, it would be interesting to investigate the two remaining cases of solutions for (\ref{Br}), namely $\frac{\lambda_r^2}{4}=8\pi G \epsilon_0$ and $\frac{\lambda_r^2}{4}>8\pi G \epsilon_0$, and analyze the how inner structure of a thick cosmic string is affected by the modification of gravity in the both cases.

\begin{acknowledgments}
The authors thank to CNPq, CAPES and FAPES for financial support.  
\end{acknowledgments}

\end{document}